\documentclass[10pt,aps,notitlepage,superscriptaddress]{revtex4-1}

\usepackage[hiresbb,final]{graphicx}
\usepackage{amsmath,amssymb,bbm}
\usepackage[export]{adjustbox}
\usepackage{graphicx,wrapfig}
\usepackage{float}
\usepackage{hyperref}
\usepackage{xcolor}

\newcommand{\sm}[1]{\begingroup\color[rgb]{0,0,0}#1\endgroup}
\newcommand{\cb}[1]{\begingroup\color[rgb]{0,0,0}#1\endgroup}

\newcommand{\Rb}{\mathbf{R}}

\newcommand{\ub}{\mathbf{u}}
\newcommand{\eb}{\mathbf{e}}

\newcommand{\Ub}{\mathbf{U}}\newcommand{\rb}{\mathbf{r}}
\newcommand{\Xb}{\mathbf{X}}

\makeatletter
\def\sgn{\mathop{\operator@font sgn}}
\makeatother

\usepackage[normalem]{ ulem } 
\usepackage{soul}  

\newcommand{\eccm}{Gulliver UMR CNRS 7083, ESPCI Paris, PSL Research University, 10 rue Vauquelin, 75005 Paris, France}

\begin{document}

\title{The flow field around a confined active droplet}

\author{Charlotte de Blois}
\author{Mathilde Reyssat}
\affiliation{\eccm}
\author{S\'ebastien Michelin}
\affiliation{LadHyX, D\'epartement de M\'ecanique, Ecole Polytechnique, CNRS, 91128 Palaiseau, France}
\author{Olivier Dauchot}
\affiliation{\eccm}

\date{\today}

\begin{abstract}
We present here first-of-a-kind experimental measurements of the flow field around a swimming water droplet, using confocal PIV in three dimensions.  
The droplet is denser than the continuous oil phase, and swims close and parallel to the bottom wall. The measured flow field is first quantitatively characterized and compared to the flow field obtained for an axisymmetric swimmer in unbounded flow. Important qualitative differences are observed, in particular the emergence of a strong isotropic radial flow field in the planes parallel to the wall. These fundamental differences stress the critical impact of confinement on the flow field around the swimming droplet. We then propose an analytical formulation, based on a reduced order description of the swimming droplet in terms of fundamental hydrodynamic singularities as well as the classical method of images. This model is able to account exactly for the effect of the wall and provides a simplified description of the flow field as the superposition of axisymmetric \cb{dipole and quadrupole} singularities. We demonstrate that this simplified description quantitatively captures the effect of the wall on the \cb{dipolar and quadrupolar components of the flow field}. \cb{Further including the confinement-induced asymmetry of the concentration field responsible for the droplet's Marangoni propulsion, our model is also able to account for the \cb{monopolar} contribution to the experimental flow field which drives the dominant far-field signature of the droplet.}
\end{abstract}


\maketitle

\section{Introduction}
\label{sec:intro}
 Achieving artificial locomotion in a surrounding fluid at the micron-scale, in order to perform a multitude of tasks in technical and medical applications, has become a central goal of nanoscience, at the interface of hydrodynamics, physico-chemical engineering and soft matter~\cite{Ebbens_2010}.
One possibility is to take inspiration from the swimming strategies adopted by micro-organisms~\cite{Lauga_2009,Hong_2010}.
As pointed out by Purcell\cite{Purcell_1977}, swimming at low Reynolds number must involve a cyclic and non-time-reciprocal motion. Examples include the rotation of bacteria's helical flagellar bundle~\cite{Lauga_2016}, the actuated motion of a sperm flagellum~\cite{friedrich_2010} or the synchronized beating of cilia on ciliated protozoa~\cite{tamm_1970}. To mimic such complex and coordinated motion is possible, as illustrated for instance by magnetically-powered microswimmers~\cite{Dreyfus_2005,Gao_2010}, but requires highly specific design of actuated multi-component systems. 

An alternative strategy consists of developing interfacial flows caused by the local interaction with the physico-chemical properties of the swimmer's immediate environment, such as the electric potential, temperature, or solute content. These interactions are responsible for surface tension gradients, also called Marangoni stresses, and/or diffusive flows within the thin interfacial region, also called phoretic flows, which can be accounted for by an apparent slip velocity~\cite{Anderson_1989}.
Examples include self-phoretic Janus swimmers, colloidal particles with asymmetric physico-chemical properties over their surface~\cite{Palacci_2010,Theurkauff_2012,Howse_2007}, and swimming droplets, the motion of which derives from a surface tension gradient~\cite{Toyota_2009,Hanczyc_2007,Thutupalli_2011,Izri_2014,Herminghaus_2014, Mino_2010,Maass_2016}. Such swimming droplets present a particular interest as they can be used for transport in micro-fluidic devices, leading to new possibilities of application~\cite{Weibel_2005,Koumalis_2013}. 

From a theoretical perspective, the steady state hydrodynamic flow exhibited by spherical microswimmers, suspended in an \emph {unbounded} medium, can be mapped onto effective hydrodynamic squirmers, the motility mechanism of which is encoded in a slip velocity prescribed at the interface of their nearly spherical body~\cite{Lighthill_1952,Blake_1971}.  This slip velocity  determines uniquely the hydrodynamic field and associated motion of spherical bodies with~\cite{Blake_1971} or without~\cite{Pak_2014} axially symmetric surface properties.

However, be it in a biological environment or in a microfluidic device, \cb{in many cases, microswimmers do not} evolve in a 3D infinite and unbounded medium~\cite{Bechinger_2016}, \sm{and several} observations indeed reveal the critical importance of confinement on the swimmer's dynamics. Several microswimmers are attracted by the boundaries~\cite{Frymier_1995,Berke_2008,Zottl_2016}, which can then be used to capture~\cite{Takagi_2014,Volpe_2011}, or steer the swimmer motion~\cite{Das_2015,Denissenko_2012,DaviesWykes_2017}. The presence of a boundary has been observed to influence not only the motion of a single particle~\cite{DiLeonardo_2011,DiLuzio_2005,Magariyama_2005} but also the collective behavior and phase transitions of swarms~\cite{Drescher_2009,Kruger_2016,Thutupalli_2018,Wioland_2013}. Ultimately, the interactions with boundaries can be used to harvest energy from the population of swimmers~\cite{DiLeonardo_2010}. 
Obtaining a reliable description of the interaction of a swimmer with a wall is thus of significant importance. It is also a first step towards a better understanding of the interactions amongst swimmers and thereby the emergence of collective behavior~\cite{Elgeti_2018,Kanso_2019}

For swimmers driven by mechanical surface distortions, it is reasonable to assume that hydrodynamic mechanisms are the dominant contributor to the motion. If the mechanical surface distortions at the origin of self-propulsion are not modified by the proximity of the wall, a  squirmer description can be used with a  prescribed and unaltered slip velocity. Even in this simplified context, solving for the exact flow around a squirmer in the presence of a wall is in general not possible. Currently, the only exact solution is that of the flow field resulting from the motion of an axisymmetric squirmer approaching a wall, along the wall normal direction~\cite{Papavassiliou_2015}. One way out consists of describing the squirmer as a linear combination of fundamental solutions to the Stokes equations and using the methods of images~\cite{Lee_1979} to compute the flow field in the presence of a wall. Such a strategy has been applied recently to the case of an axisymmetric swimmer~\cite{Spagnolie_2012}. Focusing on characterizing the accuracy of the far-field approximation, the authors show that this simplified description can be very useful, and quantitatively predictive, for describing the behavior of a selection of swimmers close to a wall.

Considering now the case of  phoretic or Marangoni swimmers, the self-generated external field, responsible for the swimming motion, is likely to be distorted by the presence of the wall, which alters the diffusion of the physico-chemical field. This was first illustrated in~\cite{Ibrahim_2015}, before it was indeed demonstrated that in the presence of boundaries the behavior of chemically active colloids is qualitatively different, even in the far field, from the one exhibited by the corresponding “effective squirmer”~\cite{Popescu_2018}. Focusing on the near-wall motion, general analytical solutions for the concentration field, velocity and rotation of the locomotor, as a function of distance and orientation of the active cap with the surface, were obtained in the form of infinite series expansions~\cite{Mozaffari_2016}. These solutions were then used to compute general trajectories and categorize the swimming regimes. Yet, for such expansions, the correspondence between each term (i.e. angular mode) to a precise set of hydrodynamic singularities of increasing order is lost, in stark contrast with the classical decomposition  of the flow field generated by a spherical swimmer in unbounded flow~\citep{Pak_2014}. 

Experimental studies of the effect of confinement on phoretic swimmers concentrate on the kinetics of the particle trajectories and very little is known about the actual flow field. On some occasions, it has been measured in the median plane of the swimmer~\cite{Drescher_2010,Thutupalli_2011,Thutupalli_2018,Campbell_2018}, and used for qualitative discussion. Yet, a precise and quantitative description of this flow field is critical, in particular to understand the role of hydrodynamic coupling between swimmers in setting their collective dynamics. To our knowledge, the three-dimensional flow field around a phoretic swimmer remains to be fully characterized experimentally, one obvious reason being that most phoretic swimmers are micron-sized particles, for which such an analysis would require truly high resolution measurements.

\begin{figure}
\includegraphics[width=\columnwidth ]{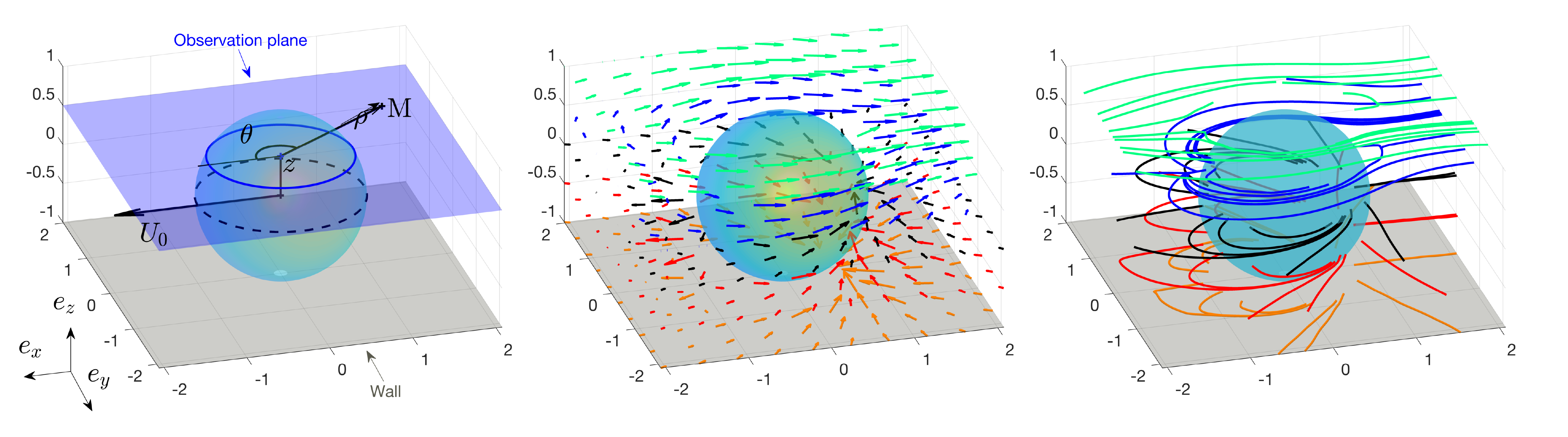}
\caption[fig:1W_3D]{{\bf Experimental flow field around a water droplet swimming close and parallel to the bottom wall of a microfluidic chamber:} Left : Sketch of the reference frames used in the present work, together with the observation planes. Middle: 3D reconstruction from the PIV analysis in planes parallel to the wall (blue plane). Right : Top view of the streamlines of the flow field. The color code indicates the height of the plane. Only a few planes are presented for clarity.} 
\label{fig:1W_3D}
\end{figure}

Here, we take advantage of large swimming water droplets~\cite{Izri_2014} to fill this gap. The water being denser than the surrounding oil medium, the droplet swims at the bottom wall of a micro-fluidic chamber. The large size of the droplet -- typically $100\mu$m in radius -- allows us to perform PIV measurements in 3D, using confocal microscopy, of the hydrodynamics flow around the swimmer (see Fig.~\ref{fig:1W_3D}). In each plane, the flow field is decomposed on a Legendre basis for the angular dependance and we observe that the flow field is well described by its decomposition on the three first modes. This allows us to fully characterize the radial dependance of the flow amplitudes along these modes. The experimental flow field is then compared with that of an axisymmetric squirmer performing a steady motion parallel to the wall, as obtained following the methodology introduced in~\cite{Spagnolie_2012}. While this simplified description quantitatively captures the radial and vertical dependence of the dipolar and quadrupolar symmetries, it completely fails at capturing the most salient feature of the experimental flow field, namely the emergence of a strong \cb{monopolar symmetry} in each plane parallel to the wall. Our results indicate that enforcing the axisymmetry of the slip velocity at the surface of the swimmer is too strong of an assumption: the wall indeed modifies significantly the concentration field responsible for the phoretic motion of the droplet, thereby breaking the axisymmetry of the problem. Including the active flows pumped by the swimmer in the vertical direction, while being held by gravity, successfully captures the \cb{monopolar symmetry of the flow field}. These results call for further analytic treatment, taking into account the effect of the wall on the concentration field. Finally, we also present and discuss the experimental flow field obtained when the droplet is confined between two parallel walls separated by typically one droplet diameter, a situation of interest in many micro-fluidic devices, for which an analytical treatment remains an open question.

The paper is organized as follows. Section~\ref{sec:setup} describes the experimental system, recalling the swimming mechanism of the droplet and describing the PIV methods and data processing techniques -- full details are provided in the Appendix. Section~\ref{sec:1W} synthesizes our main experimental findings regarding the flow field around a swimming droplet performing steady motion close and parallel to a wall~(Sec.~\ref{sec:1W-exp}) and its comparison to a simplified theoretical prediction in terms of an axisymmetric swimmer~(Sec.~\ref{sec:1W-theo}). Finally Section~\ref{sec:2W-exp} describes the experimental flow around a droplet swimming between two confining parallel walls.

\section{Setting}
\label{sec:setup}
The experimental system consists of water droplets, with diameter $a\simeq 100\mu$m, swimming in a continuous oil-surfactant phase consisting of a mono-olein solution, a nonionic surfactant, with concentration far above the critical micellar concentration. The swimming motion of the water droplets results from the combination of two ingredients~\cite{Izri_2014}. The first is intimately linked to the system being away from its physico-chemical equilibrium: at room temperature, the stable thermodynamic phase of the system is a micro-emulsion with all inverse micelles in the oil-surfactant phase filled with water. As a result, a slow but steady flux of water takes place from the droplet to the inverse micelles. In an unbounded medium, this in turn leads to the development of radial concentration fields of empty and swollen micelles. The second ingredient is that this isotropic concentration field is unstable against an infinitesimal flow disturbance. Suppose the droplet experiences a tiny displacement; the concentration field around the droplet is no longer isotropic and concentration gradients parallel to the droplet surface appear. These gradients in turn induce Marangoni stresses and phoretic flows. The resulting mobility is such that the droplet moves further towards the regions of small concentration and the initial disturbance is thus amplified. For this instability to take place, the P\'eclet number $Pe=U^* a/D$, where $U^*$ is the characteristic swimming speed, and $D$ is the diffusion coefficient of the micelles, must exceed some critical value $Pe_c=O(1)$. In other words, the diffusion of the micelles must be slow enough as compared to their advection by the Marangoni and phoretic flows. This instability mechanism first described in the context of isotropic phoretic colloids~\cite{Michelin_2013} was then generalized to the case of droplets, for which Maragoni stresses are also present~\cite{Izri_2014}.  

The droplets are produced using a \copyright Femtojet apparatus by injecting a single droplet of controlled size in a circular micro-fluidic chamber of diameter $1$ cm.
The droplets have a diameter $a \simeq 100 \mu$m. They are filled with a (milli-Q)  water solution of $15$\%wt NaCl. The continuous phase is a $25$ mM mono-oleine surfactant (MO; 1-oleoylrac-glycerol, $99$\%, Sigma-Aldrich) solution in squalane (Sq; 99\%, Sigma-Aldrich). The room temperature is kept above $25^{o}C$ in order to avoid mono-oleine crystallization.
The observation chambers are made up of an UV-curing glue (Norland Optical Adhesive No. 81, NOA). In the so-called one-wall geometry, the chamber has height $2h\gtrsim 5$ mm $\gg a$ and is left opened to the air. In this geometry, the droplet swims at the bottom of the chamber, far from the top free surface since the water phase is denser than the oil phase ($d_{Sq}=0.8$). In the so-called two-wall geometry, the chamber has height $2h\simeq a$ and is closed on its top by a NOA coated glass cover slip. In this geometry, the droplet swims confined between two walls. \cb{In both cases, the droplet, the diameter of which does not exceed the capillary length, remains spherical.} 

The droplet starts swimming immediately and reaches steady motion after a few minutes, with a constant velocity $V_0\simeq 20 \mu$m/s, following a trajectory with a typical persistence length of a few droplet diameters.
The droplet motion parallel to the wall is tracked in the frame of the laboratory, where $(x,y)$ denote the coordinates parallel to the bottom wall and $z$ the normal coordinate. The origin of the $z$-axis is located at the center of the droplet. In the following, the droplet radius $a/2$ and velocity $U_0$ are used as characteristic length and velocity, so that $z_{wall} = -1$ denotes the position of the bottom wall.

The flow field around the droplet is then measured using PIV. Red fluorescent colloids tracers (Fluoro-Max$^{TM}$, 0.6 $\mu$m Red Fluorescent Polymer Microspheres, Thermo scientific) are added in the oil phase and the flow field is acquired with a CCD camera (Andor Zyla 5.5) in successive planes parallel to the wall, separated by $5\pm 1 \mu$m using confocal microscopy with a x10 objective. The acquisition frequency is $10$ frames/s and the exposure time is $50$ ms. The spatial resolution parallel to the wall is $0.65\mu$m/pixel. \cb{The relative error related to the Brownian motion of the tracers is smaller than $10^{-4}$ and can be safely ignored.}

For each experiment, $50$ images of the droplet and the surrounding flow field are acquired at each $z$. 
The PIV analysis, performed using the PIV\_lab~\cite{Thielicke_2014} code on \copyright Matlab in each $z$-plane, provides us with the velocity field in cartesian coordinates attached to the lab frame at each time step (see Appendix~\ref{App:Meth}). Note that we don't have access to the $z$ component of the velocity. In principle, this component could be deduced using mass conservation, but the method is highly sensitive to experimental noise and we here choose to restrict ourselves to the analysis of the velocity components parallel to the bottom wall.

We focus exclusively on trajectories (i) where the droplet is far from the lateral boundaries of the observation chamber and (ii) during time windows corresponding to steady motion of the droplet along linear trajectory. This allows us to average the instantaneous flow fields obtained from PIV, at each $z$, thereby reducing the experimental noise. \\

We are now in position to describe quantitatively the reconstructed flow field in the one-wall geometry as illustrated on Fig.~\ref{fig:1W_3D}. 
The first step is to adopt a suitable system of coordinates. While spherical coordinates centered on the droplet are natural to describe an axisymmetric swimmer in an unbounded domain, the presence of the wall here calls for a description using cylindrical coordinates ($\rho$,$\theta$,$z$) with the $z$-axis orthogonal to the wall and $\theta=0$ the swimming direction.
Recalling that we use the droplet radius as the unit length, $\rho=1$ describes a cylinder around the droplet, tangent to the median plane $(z=0)$. Additionally, for a steady and linear motion, the flow field conserves a planar symmetry with respect to the vertical plane $\theta=0$. Exploiting this parity symmetry, the radial and azimuthal components of the dimensionless velocity field in each plane are decomposed onto the basis of Legendre polynomials: 

\begin{align}
\label{eq:Legendre_rho}
u_\rho(\rho,\theta,z)&=\sum_{n=0} \phi_\rho^n(\rho,z) L_n(\nu) \\
u_\theta(\rho,\theta,z)&=\sum_{n=1} \phi_\theta^n(\rho,z) L_n^1(\nu),
\label{eq:Legendre_theta}
\end{align}
where $\nu=\cos{\theta}$. Note that the flow field is measured in the lab reference frame and the velocities are zero far from the droplet.  $L_n(\nu)$  (resp.  $L_n^1(\nu)=-\sqrt{1-\nu^2}L_n'(\nu)$) are the Legendre polynomials (resp.  associated Legendre polynomials of the first kind). 
These Legendre polynomials form an orthogonal basis of hydrodynamic \cb{azimuthal symmetries} with  $\phi_{i}^n(\rho,z)$ the dimensionless amplitude of the $n^{th}$-\cb{multipolar symmetry} for the radial $(i=\rho)$ and azimuthal $(i=\theta)$ components, which are obtained by the following projections: 
\begin{align}
\label{eq:proj_rho}
\phi^n_\rho(\rho,z)&=\frac{2 n+1}{2}\int_{-1}^1 u_\rho(\rho,\theta,z) L_n(\nu) d\nu,\\
\phi^n_\theta(\rho,z)&=\frac{2 n+1}{2 n (n+1)}\int_{-1}^1 u_\theta(\rho,\theta,z) L_n^1(\nu) d\nu.
\label{eq:proj_theta}
\end{align}

\section{Swimming close to a wall}
\label{sec:1W}

\subsection{Experimental flow field}
\label{sec:1W-exp}

\begin{figure}[t!]
\includegraphics[width=0.7\columnwidth ]{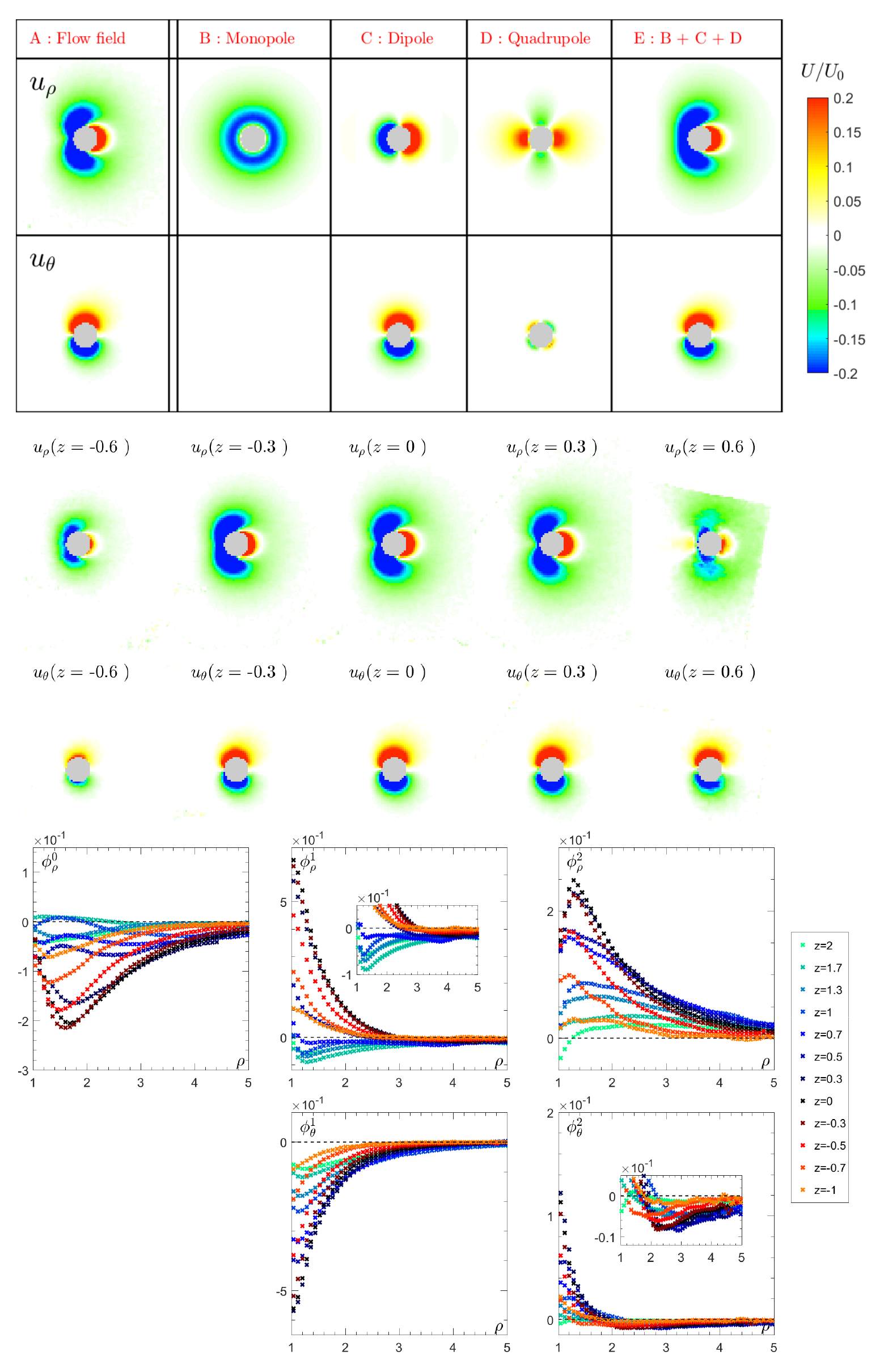}
\caption{{\bf Experimental flow field in the one-wall geometry --} Top: $u_\rho$ and $u_\theta$ in the median plane ($z=0$). The complete experimental flow field (A) is shown as well as \cb{its monopolar component (B), its dipolar component (C) and its quadrupolar component (D)}. The flow reconstruction from the first three components (i.e. the sum of the contributions in B, C and D) is also shown (E).  Middle: $u_\rho$ (top) and $u_\theta$ (bottom) for $z=-0,6$, $z=-0,3$, $z=0$; $z=0,3$ ; $z=0,6$ (same color code as above). Bottom: Amplitudes $\phi^n_\rho(\rho,z)$ and $\phi^n_\theta(\rho,z)$, of the first three modes $(n=0,1,2)$ as a function of the radial coordinate $\rho$ for different $z$.}
\label{fig:1W}
\end{figure}

We consider here the one-wall geometry with the water droplet performing a steady linear motion at the bottom of the chamber, parallel to the wall. Our experimental results are summarized on Fig.~\ref{fig:1W}. 

Let us first concentrate on the flow field in the median plane $(z=0)$.
The top panel of Fig.~\ref{fig:1W} displays color coded maps of the radial and azimuthal components of the flow field, together with its decomposition onto the three first \cb{hydrodynamic multipolar components with respectively monopolar, dipolar, and quadrupolar symmetry, and the reconstruction of the flow field from only these three first components.}
One observes that the reconstructed flow field is very similar to the original one. In the following, we shall therefore restrict our analysis to the first three modes $(n=0,1,2)$. 
Second, and most remarkably, the radial component of the flow field exhibits a strong \cb{monopolar symmetry}, which actually dominates at long range. In the median plane $z=0$, the cylindrical and spherical coordinates  description of the flow field are strictly equivalent, and it is well known that the flow field around an axisymmetric swimmer in an unbounded medium includes no \cb{monopolar symmetry contribution}: our observations therefore provide a strong indication of the influence of the wall on the flow field.

The influence of the wall is further characterized by considering the flow field components in different planes parallel to the wall (see the middle panel of Fig.~\ref{fig:1W}). The flow field appears strongly asymmetric with respect to the median plane, as could already be noticed from Fig.~\ref{fig:1W_3D}. This is particularly true for the radial component $u_\rho$, which has a nearly \cb{dipolar symmetry} close to the wall and a nearly quadrupolar one when approaching the top of the droplet. The azimuthal component $u_\theta$ conserves the same dipolar symmetry across the droplet height, but one still notices a faster decrease of the flow intensity away from the droplet in planes closer to the wall.
The complete quantitative description of the flow field is finally provided on the bottom panel of Fig.~\ref{fig:1W}, where the amplitudes of the first three azimuthal modes are displayed as a function of the radial distance to the droplet $\rho$ and the distance to the wall $z$. 

In order to fully appreciate the effect of the wall, one shall compare these profiles with those of a model swimmer in an unbounded medium. Solving the Stokes equation, one obtains the velocity field around an axisymmetric swimmer as an infinite sum of  hydrodynamics singularities, for any given slip velocity $\ub_s$ at the interface~\cite{Blake_1971,Lighthill_1952}. In agreement with our experimental observations, we limit our description of the flow field to the singularities, which contribute to 
flow fields with azimuthal symmetries up to the quadrupolar order. The swimmer is thus described by the most commonly used squirmer model, namely the superposition of a \cb{Stokes (force) dipole}, $\ub_{fd}(\rb)\sim 1/r^2$, a source dipole, $\ub_{sd}(\rb)\sim 1/r^3$ and a source quadrupole $\ub_{sq}(\rb)\sim 1/r^4$. In unbounded conditions, this swimmer generates the flow field:
\begin{equation}
\ub(\rb) = \lambda \ub_{sd}(\rb) + \zeta \ub_{sq}(\rb) + \kappa \ub_{fd}(\rb),
\label{eq:unbounded_squirmer}
\end{equation}
where the dimensionless coefficients $\lambda, \zeta$ and $\kappa$ are fixed by the slip velocity prescribed at the swimmer boundary $\ub(\rb)=\Ub_0+\ub_s$ :  one finds $\lambda = 0.5$ and $\kappa = -3\zeta$.
The flow field generated by source and force multipoles are successively obtained from the flow field generated by a singular point source and point force. In particular $\ub_{sd}(\rb)=-\nabla \ub_{s}(\rb)\cdot\eb$, $\ub_{sq}(\rb)=-\nabla \ub_{sd}(\rb)\cdot\eb$ and $\ub_{fd}(\rb)=-\nabla \ub_{f}(\rb)\cdot\eb$, with $\eb$ the unit vector pointing in the swimming direction. The corresponding amplitude profiles are then obtained by expressing the flow field in the cylindrical system of coordinates $(\rho,\theta,z)$ and projecting it on the Legendre polynomials, Eqs.~\eqref{eq:proj_rho}--\eqref{eq:proj_theta}.
The vectorial expression of the flow fields and the analytical expressions of the amplitudes are provided in Eqs.~\eqref{eq:0W-vect-1}--\eqref{eq:0W-vect-2} and \eqref{eq:0W-amp-1}--\eqref{eq:0W-amp-2}, together with their graphical representation (Figure~\ref{fig:0W_theo}). It should be noted that the source quadrupole and the Stokes dipole, which in spherical coordinates only contribute to the quadrupolar symmetry of the flow, here also contribute to a monopolar symmetry for the radial and azimuthal velocity components $u_\rho(\rho,\theta)$ and $u_\theta(\rho,\theta)$ when out of the median plane. 

A number of peculiar features in the radial dependence of the amplitudes should be noted when comparing the experimental amplitude profiles, Fig.~\ref{fig:1W}, with the ones obtained for the model axisymmetric squirmer in an unbounded geometry (i.e. without taking into account the influence of the wall) as shown on Fig.~\ref{fig:0W_theo}.
While for the unbounded squirmer, the amplitude $\phi^0_\rho$ always presents a minimum, we observe in the experimental data that this minimum turns into a maximum when moving towards the upper hemisphere of the droplet. Similarly, the experimental amplitude $\phi^1_\rho$ displays a negative minimum in the upper part of the droplet, where the squirmer model only presents a positive maximum; experimentally, $\phi^1_\theta$ further presents a minimum in the upper part of the droplet while this quantity is monotonously increasing in the squirmer case. Finally, regarding the quadrupolar symmetry, the experimental measurements for $\phi^2_\rho$ appear essentially similar to that of the unbounded squirmer, but $\phi^2_\theta$ presents a systematic negative minimum for all $z$ in the droplet case, while it is monotonically decreasing and positive in the case of the unbounded squirmer.
Altogether, one sees that the flow field around the droplet is, as expected, strongly affected by the presence of the wall, and is therefore poorly accounted for  by an unbounded squirmer model. It is also by far more complex than the naive intuition one could develop from an observation limited to the median plane.

\subsection{Comparison with a squirmer model moving parallel to a wall}
\label{sec:1W-theo}
Capturing theoretically the complex flow fields reported experimentally is highly non trivial. A possible strategy would consist in solving exactly, up to numerical truncations, for the hydrodynamics and concentration fields around the droplet in the presence of the wall~\cite{Mozaffari_2016}. This however requires the knowledge of the boundary conditions at the boundary of the swimming body. For a Janus colloid, with an active cap of a prescribed geometry, it is already a strong hypothesis to assume that the phoretic slip velocity is not altered by the wall-modified concentration field. In the present case, we recall that the phoretic slip velocity and Marangoni stresses both result from a linear instability. In other words, in the bifurcated non-linear solution describing the swimming motion, they are functions of both the concentration  and  hydrodynamic fields. The interaction with the wall thus becomes a highly nonlinear problem, which cannot be solved easily following this strategy.

In the absence of exact treatment, it is still desirable to know how far a simplified model may account for the experimental description. With this goal in mind, we propose here an alternate and approximate model as an axisymmetric squirmer moving parallel to the wall, and computing the associated flow field using the method of images~\cite{Spagnolie_2012}. We stress that this description is simplified in the sense that (i) it overlooks the dynamics of the concentration field and its impact on the hydrodynamic boundary condition at the swimmer's surface; (ii) it is a far-field approximation and therefore overlooks that the droplet radius and distance to the wall are comparable. Whether it would be able to capture the complex structure of the flow field reported experimentally is therefore far from obvious \emph{a priori}. 

We shall see that it is in fact able to describe surprisingly well the \cb{quadrupolar and dipolar components} but fails to explain the emergence of a strong \cb{monopolar symmetry} for the components of the flow field parallel to the wall. \\

The squirmer model considered is that introduced in the previous section. The presence of a no-slip infinite plane wall imposes a vanishing flow velocity $\ub={\bf 0}$ at the wall. The methods of images followed here consists of introducing singularities at the mirror position of the swimmer with respect to the wall, such that the flow field, obtained from the superposition of the original and image singularities, satisfies the no slip condition at the wall exactly. Note that the hydrodynamic image systems differ from the simple mirror image of the original singularity, as is the case in other fields such as diffusion or electrostatics, where the field satisfies Laplace's equation.
\begin{figure}[b!]
\includegraphics[width=0.715\columnwidth ]{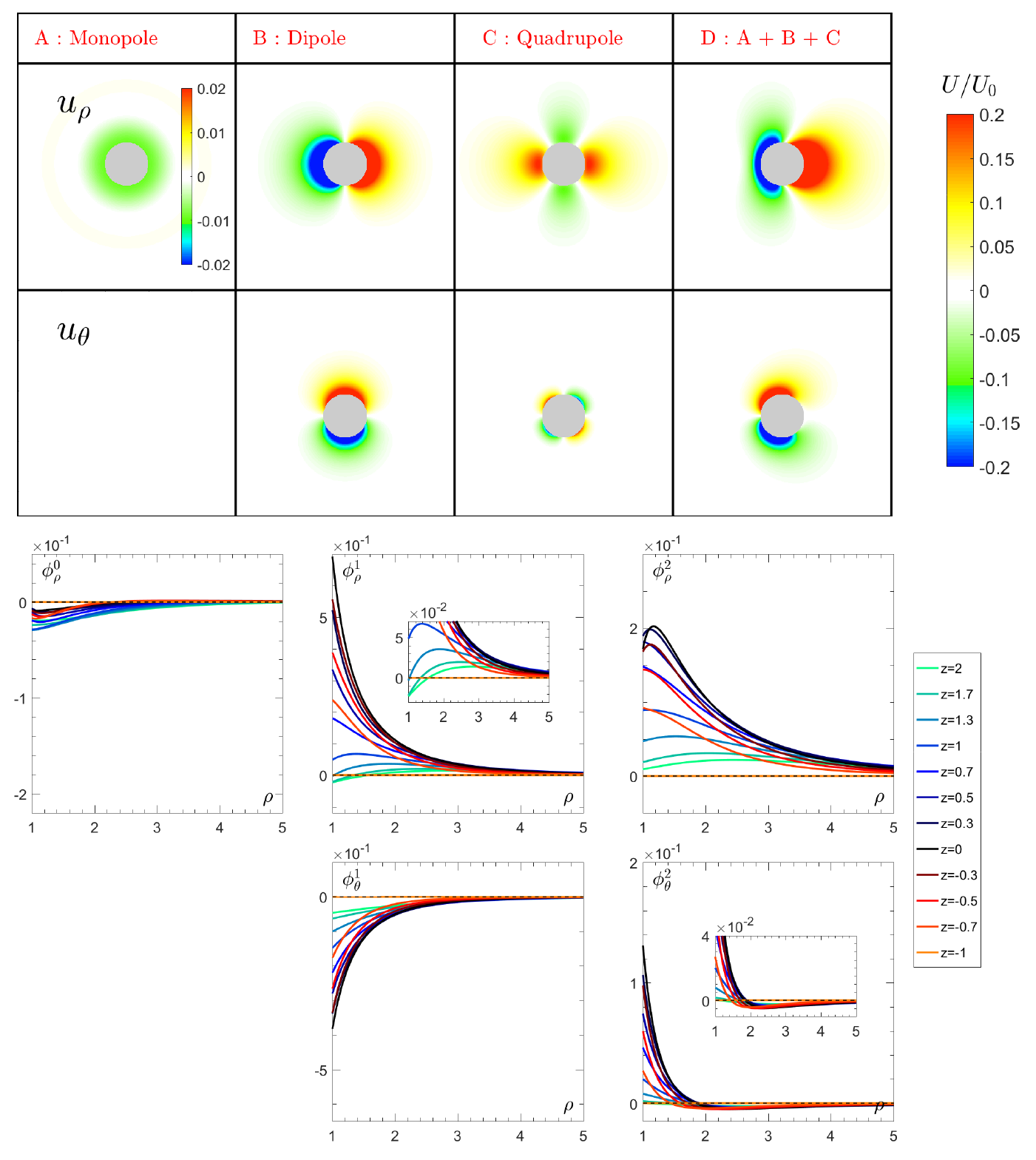}
\caption{{\bf Model flow field in the one-wall geometry:} Top: $u_\rho$ and $u_\theta$ in the median plane, computed from the first singularities and their image systems, Eq.~\eqref{eq:confined_squirmer}, using the value of $\kappa$, $\lambda$ and $\zeta$ obtained from the experimental data. \cb{A: Monopolar component generated by the original Stokes dipole and the original source quadrupole. B: Dipolar component generated by the source dipole C: Quadrupolar component generated by the stokes dipole and the source quadrupole} D: Total flow field generated by the first singularities : sum of A, B and C. The color-scale codes for the amplitude of the velocity. Bottom: Corresponding amplitudes $\phi^n_\rho(\rho,z)$ and $\phi^n_\theta(\rho,z)$, of the first three modes $(n=0,1,2)$ as a function of the radial coordinate $\rho$ for different $z$.}
\label{fig:1WT}
\end{figure}

The appropriate singularities have been identified for a source monopole, a source dipole or a point force in the classical work of Blake and Chwang~\cite{Blake_1974}. Spagnolie and Lauga~\cite{Spagnolie_2012} recently computed the image systems of higher order singularities including the stokes dipole. The image flow field for the source quadrupole can be obtained similarly. 
The general method consists again of using the fact that derivatives of the source or force singularities produce other higher-order singularity solutions of the Stokes equations: deriving the known images of lowest order singularities, one obtains the images of the higher order singularities of interest $\ub^*_{sd}(\rb), \ub^*_{sq}(\rb)$ and $\ub^*_{fd}(\rb)$. The flow field in the presence of the wall then simply reads:
\begin{equation}
\ub^*(\rb) = \lambda \ub^*_{sd}(\rb) + \zeta \ub^*_{sq}(\rb) + \kappa \ub^*_{fd}(\rb).\label{eq:confined_squirmer}
\end{equation}
The vectorial expressions of the so-obtained flow fields are provided in Appendix~\ref{App:Theo1W}, Eqs.~\eqref{eq:1W-usd}--\eqref{eq:1W-ufd}.  Note that these solutions of the Stokes equations do not satisfy the boundary conditions at the interface of the droplet, so that $\lambda$, $\kappa$ and $\zeta$ are now unknown -- thus acknowledging the fact that the surfactant-induced velocity condition at the interface of the droplet is modified by the presence of the wall. This flow field can then be expressed in the cylindrical coordinate system and projected onto the Legendre basis (see Eqs.~\eqref{eq:1W-amp-1}--\eqref{eq:1W-amp-2}). 

In order to compare quantitatively the experimental flow fields and the predictions of the above squirmer model, the coefficients $\lambda, \zeta$ and $\kappa$ are extracted from the experimental data as follows. The \cb{dipolar symmetry of the flow field} arises from the source  dipole only, \cb{while the monopolar and quadrupolar symmetries of the flow field result from both the stokes dipole and source quadrupole}:
$\lambda$ is thus obtained by minimizing  $\left<\parallel\!\!(\Delta\phi_\rho^1)^2\!\!\parallel+\parallel\!\!(\Delta\phi_\theta^1)^2\!\!\parallel\right>$, where $\left<\parallel\!\!\Delta\phi\!\!\parallel\right>$ denotes the average over the experimental realizations of the $L_2$-norm of the difference between the amplitudes $\phi(\rho,z)$ measured experimentally and computed analytically.
Similarly $\kappa$ and $\zeta$ are obtained by minimizing $\left<\parallel\!\!(\Delta\phi_\rho^2)^2\!\!\parallel+\parallel\!\!(\Delta\phi_\theta^2)^2\!\!\parallel\right>$. We thereby obtain $\lambda=0.35$, $\zeta=-0.08$ and $\kappa=0.34$. This method is quite robust as for repeating the experiment for different droplets of radius of $60$ and $70 \mu$m, we obtain very similar values with interval $\pm0.05$.  The flow fields and profiles are displayed on Fig.~\ref{fig:1WT}.

A remarkable feature is that the images of the first order singularities do induce a \cb{monopolar component} even in the median plane. This \cb{monopolar component} is however much weaker than in the experimental case, and does not contribute significantly to the reconstructed flow field at odd with the experimental observation. A closer look at the amplitude profiles on Fig.~\ref{fig:1WT} (bottom) also reveals that the \cb{monopolar component does} not have the proper dependance in $z$: \cb{the monopolar symmetry of the experimental flow is more pronounced} in the median plane, while it is found in this model to be dominant above the droplet.

\begin{figure}[t!]
\includegraphics[width=0.75\columnwidth ]{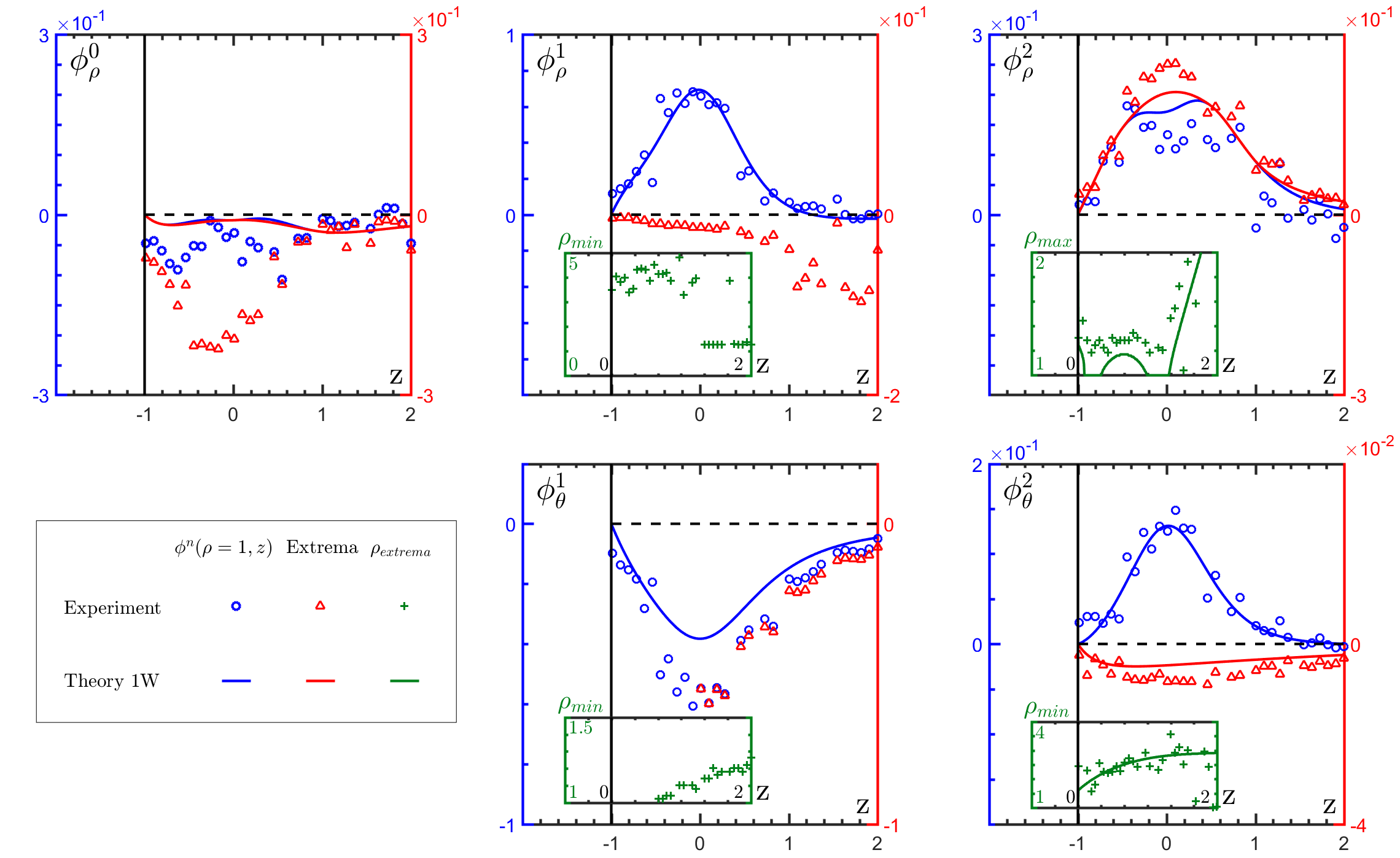}
\caption[fig:1W_max]{{\bf Comparison between the characteristics of the experimental (dots) and theoretical (straight lines) velocities in the one-wall geometry:} Evolution with $z$ of the amplitude at $\rho=1$ (blue), of the extremum (red) and of the extremum's position (green) \cb{of the monopolar component (left), the dipolar component (middle) and the quadrupolar component (right) of the velocities $u_\rho$ (top) and $u_\theta$ (bottom)}.} 
\label{fig:1W_max} 
\end{figure}

In contrast, the \cb{dipolar and quadrupolar components} of the experimentally-measured flow (Figure~\ref{fig:1W}) are surprisingly well described by this simple model (Figure~\ref{fig:1WT}), especially for the latter. The asymmetry of the flow with respect to the median plane is properly captured, and  the radial dependence of the amplitudes are also well described. This is particularly well illustrated by the non-trivial dependence on $z$ of the amplitude and position of the maximum of $\phi^2_{\rho}(\rho,z)$ and of the minimum in $\phi^2_{\theta}(\rho,z)$ (Figure~\ref{fig:1W_max}).
As far as the quadrupolar symmetry is concerned, the model predictions, together with the extraction of the parameters $\zeta$ and $\kappa$ from the experimental data, perfectly describes the experimental flow field.
More specifically $\phi_\rho^2$ has a local maximum, the position of which, varies in a non trivial way with $z$, which is well captured by the model. Similarly, the amplitude and position of the local minimum of $\phi_\theta^2$ are well reproduced.
\cb{The amplitudes of the dipolar component} are also well captured, despite some exceptions. The magnitude of the azimuthal amplitude $\phi_\theta^1$ is slightly underestimated by the model. Also,  experimentally, $\phi_\rho^1$ presents a local minimum that becomes negative close to $\rho=1$ and for $z>0.7$, while the model predicts a negative value at $\rho=1$ for $z>1$, but no local minimum. Finally, $\phi_\theta^1$ presents a local maximum close to $\rho=1$ in experiments, that is not predicted by the theory.\\

In summary, the axisymmetric squirmer model presented above predicts very well the behavior of the \cb{quadrupolar symmetry of the experimental flow field, captures the amplitude of the dipolar components but not the details of the amplitude profiles very close to the droplet, and fails to describe the monopolar symmetry observed experimentally}. Given the oversimplified nature of the model, it is already amazing that so many key features of the real flow are captured. For instance the fact that the amplitude profiles very close to $\rho=1$ are not so well captured can easily be understood given that the method of images does not guarantee the boundary conditions at the swimmer interface.  
The main limitation of the present description lies in its failure to describe properly the \cb{monopolar symmetry} of the flow observed in the experiments. This is all the more problematic since this contribution to the flow field is also observed to dominate at long range and therefore is expected to control the interactions with other swimmers: when this monopolar symmetry is present, it dominates the stokes dipole signature of the droplet, thereby rendering the common description of microswimmers in terms of pusher/ puller inadequate.

\subsection{Origin of the monopolar symmetry of the flow}
It should first be stressed that the monopolar symmetry of the flow cannot be explained by including higher order terms of the multipole expansion describing axisymmetric swimmers: although such terms do indeed contribute to a monopolar symmetry of the flow when expressed in cylindrical coordinates, their radial dependence decreases faster and faster with $\rho$ and is therefore unable to explain a dominant monopolar symmetry at large distance.  
One must therefore look for the origin of the monopolar symmetry of the flow  in low-order singularities. The point source singularity is the most natural candidate, especially given that the swimming mechanism of the droplet involves a water flux through the swelling of the micelles in the oil phase. This flux is however of microscopic nature and much too weak to account for a significant hydrodynamic flow.

To make further progress, one should realize that the slip velocity at the swimmer boundary is unlikely to remain axisymmetric with respect to the direction of motion, as implicitly assumed in the squirmer description presented above. This can be understood easily in the case of phoretic swimmers, as the concentration gradients at the surface (and resulting slip velocity) are likely altered by the presence of the wall. In the present case, where the swimming mechanism results from the non-linear advective coupling of the concentration  and hydrodynamic fields, one expects an even stronger effect of the wall. 

\begin{figure}[b!]
\includegraphics[width=0.7\columnwidth ]{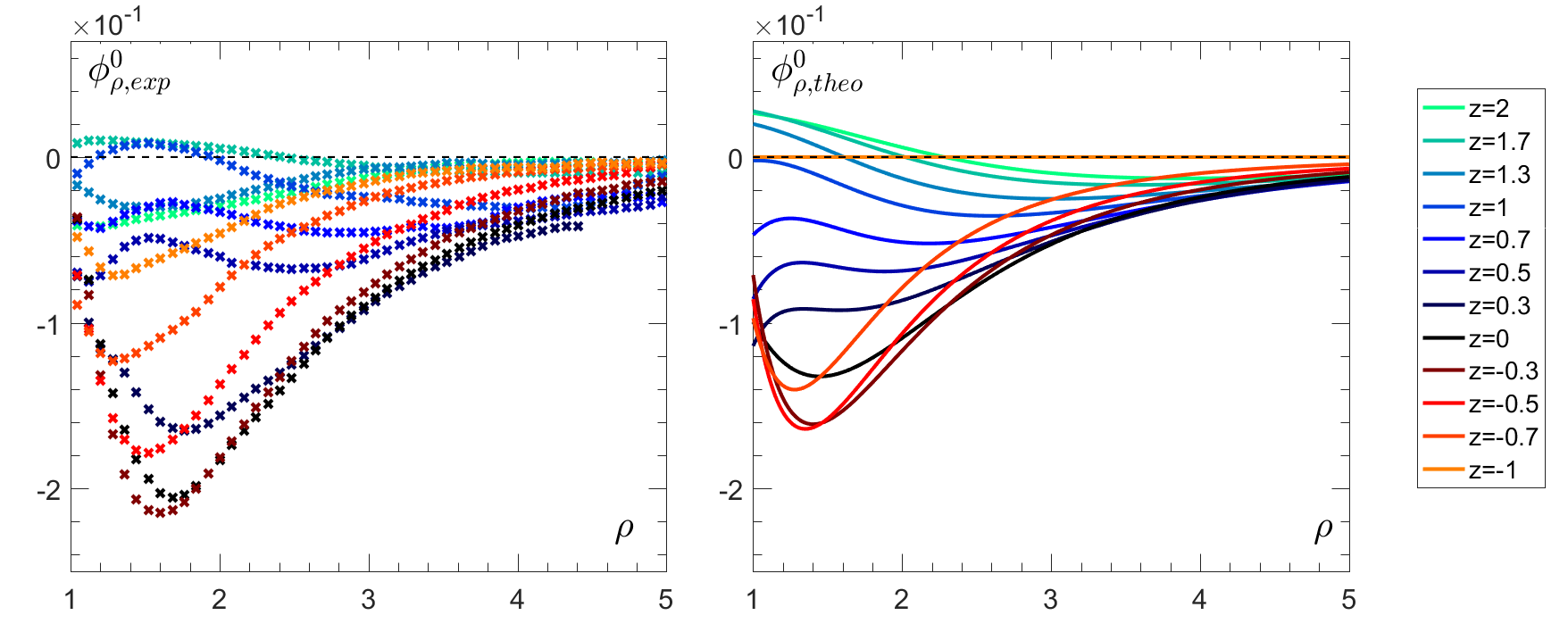}
\caption[fig:1W_G]{\cb{{\bf The monopolar symmetry :} Amplitude of the monopolar component} of the flow field measured experimental (left), and computed in the case of a non axisymmetric squirmer close to a wall, for which phoretic flows develop perpendicular to the wall, while the swimmer is held by gravity (right).} 
\label{fig:1W_G} 
\end{figure}

The full description of the coupled hydrodynamics and concentration fields, when swimming close to a wall is beyond the scope of the present work. Yet, as a first step in this direction, we may include a first correction to the axisymmetric assumption resulting from an increase of the concentration near the wall, simply because of confinement between the wall and droplet surface~\cite{Michelin_2015}. This in turn generates vertical concentration gradients along the surface of the droplet, promoting its motion away from the wall~\cite{Yariv_2016}. The droplet being denser than the surrounding fluid, it is held vertically in place by buoyancy effects, and acts as a phoretic pump in the direction orthogonal to the wall. In this case it is no longer force-free, and a vertical point force should therefore be added. The key point to notice here is that this singularity gives rise to a \cb{flow with a dipolar symmetry} in the direction orthogonal to the wall but displays a \cb{monopolar symmetry} if observed in planes \emph{parallel} to the swimming direction. We also notice that the amplitude profiles of the \cb{monopolar components} are non-monotonous. This suggests that more than one singularity should be considered. We thus also include the source dipole perpendicular to the wall.

The flow fields produced by the vertical point force and source dipole in the presence of the wall are again computed using the methods of images, and expressed in cylindrical coordinates (see Appendix~\ref{App:TheoPerp}).  The contribution of these two singularities in planes parallel to the wall \cb{has a purely monopolar symmetry}, so that they only contribute to the amplitude $\phi_\rho^0(\rho,\theta)$, without altering the higher-order angular amplitudes.  
\cb{The amplitude of the monopolar component} $\phi_\rho^0(\rho,\theta)$ now depend on four coefficients: $\zeta, \kappa, \epsilon$ and $\sigma$ the magnitudes of the parallel source quadrupole, $\ub^*_{sq}$, the parallel stokes dipole $\ub^*_{fd}$, the perpendicular point force $\ub^*_{pf,\perp}$ and the perpendicular source dipole $\ub^*_{sd,\perp}$, respectively. $\kappa$ and $\zeta$ have already been determined from the experimental data using the higher order singularities, and thus equal to the values computed in the previous section. $\epsilon$ and $\sigma$ are obtained by minimizing $\left<\parallel\!\!(\Delta\phi_\rho^0)^2\!\!\parallel\right>$. The resulting amplitude profiles are provided on Fig.~\ref{fig:1W_G}, together with the experimental profiles. Including the contribution of the vertical singularities resulting from confinement of the concentration field is observed to capture very well the key trends observed experimentally. 

\section{Swimming between two walls}
\label{sec:2W-exp}
We finally consider the two-wall geometry with the water droplet performing a steady linear motion between a top and a bottom wall separated by typically one droplet diameter. In this case one cannot use the method of images as it would result in an infinite set of images with respect to the two walls. Such a method is only tractable when looking at the flow field far from the droplet in the horizontal direction~\cite{Liron_1976,Thutupalli_2018,Kanso_2019}. In this double-confinement geometry, there is therefore little hope to derive analytical expressions for the flow field close to the droplet even within very simple approximations. Yet, as we shall see below, it remains of interest to discuss the qualitative difference with the case of one wall confinement, from the purely experimental point of view. 

\begin{figure}[t!]
\includegraphics[width=0.715\columnwidth ]{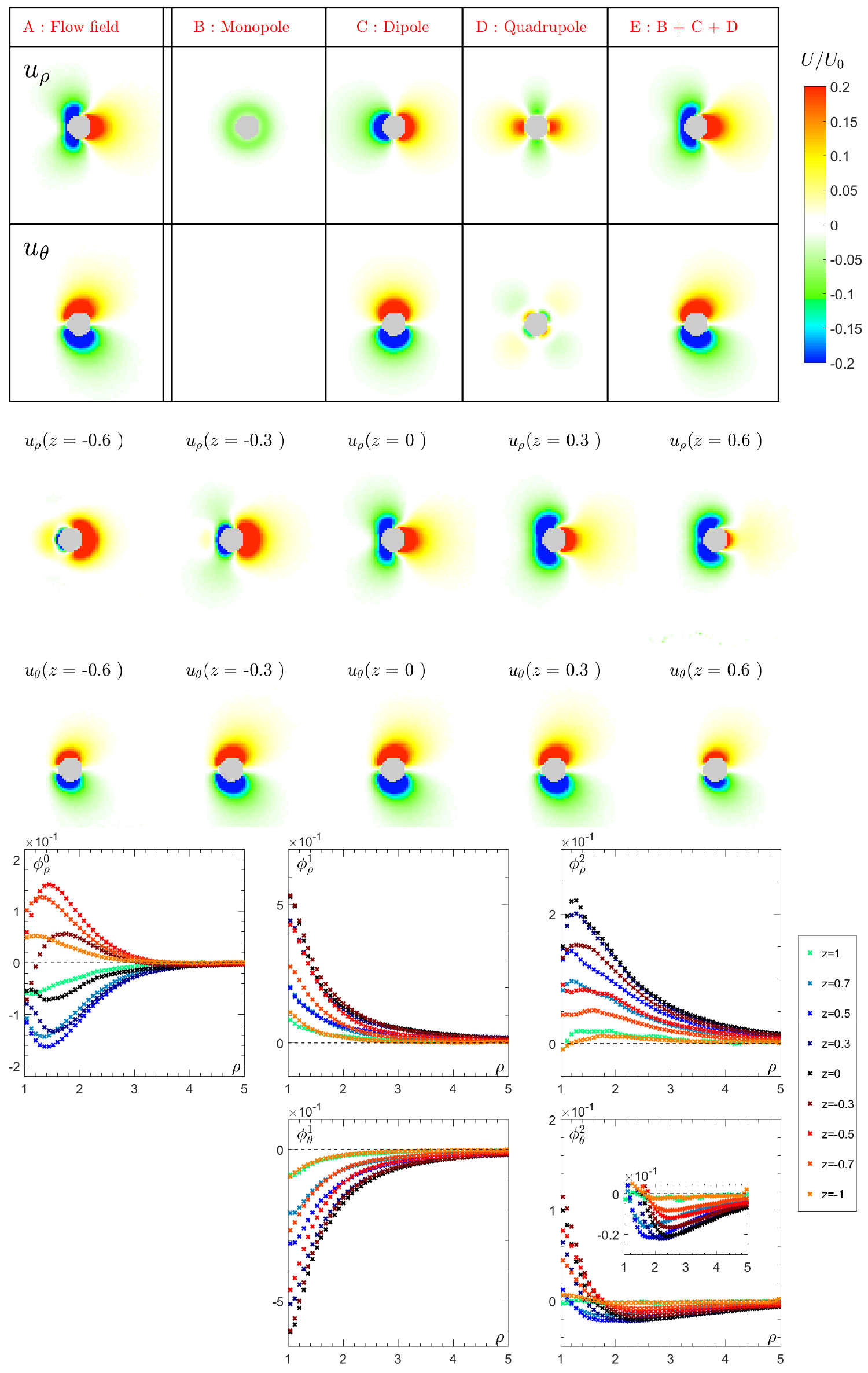}
\caption{{\bf Experimental flow field in the two-walls geometry:} Top: $u_\rho$ and $u_\theta$ in the median plane. A. Experimental flow field. B. \cb{Monopolar component} i.e projection onto $L_0$. C. \cb{Dipolar component}, i.e. projection onto $L_1$ and $L^1_1$. D. \cb{Quadrupolar component}, i.e. projection onto $L_2$ and $L^1_2$. E. Reconstruction from only the three first components, i.e. sum of B, C and D. Middle: $u\rho$ (top) and $u_\theta$ (bottom) for $z=-0.6$, $z=-0.3$, $z=0$; $z=0.3$ ; $z=0.6$ (same color code as above). Bottom: Amplitudes $\phi^n_\rho(\rho,z)$ and $\phi^n_\theta(\rho,z)$, of the first three modes $(n=0,1,2)$ as a function of the radial coordinate $\rho$ for different $z$.}
\label{fig:2W}
\end{figure}

The experimental results are summarized on Fig.~\ref{fig:2W}, in the same way as for the one wall geometry. We first note (top panel) that, here also, the flow field reconstructed from the decomposition on the first three hydrodynamic multipoles conveys all the experimental signal. 
As in the one-wall case, the radial component of the flow field \cb{presents a monopolar symmetry}. However, this \cb{monopolar component} is here small compared to the one-wall case, and does not dominate at long range.  A natural explanation is that the effect of the vertical force singularity, which dominates at large distance in the one-wall case, is exponentially screened in the far-field in double confinement~\cite{Liron_1976}. The contribution to the flow field are thus mainly \cb{dipolar and quadrupolar symmetries}. 

The influence of the wall is further characterized by considering the flow field symmetries, observed at different distances from the wall (see the middle panel of Fig.~\ref{fig:2W}). The main observation is that the flow field is strongly asymmetric with respect to the median plane \cb{(asymmetry in $z$)}, despite the approximate top-down symmetry of the problem.
This asymmetry is however not found on all modes. This is best observed on the amplitude profiles displayed on the bottom panel. In the case of the radial component, \cb{the dipolar and quadrupolar symmetries dominate the median plane flow, while on the contrary the monopolar symmetry dominates the flow close to the walls. This monopolar component} is positive at the bottom wall and negative at the top wall. It is thus the main contributor to the flow field asymmetry. 
\cb{The azimuthal component, dominated by a dipolar symmetry, does not present such an asymmetry in $z$. There is an asymmetry carried by the  azimuthal quadrupolar symmetry}, but the latter being very weak, it does not affect qualitatively the flow field.
Altogether the \cb{monopolar symmetry} is by itself responsible for the observed asymmetry of the flow field. Since we never observe the reverse solution with a positive, respectively negative, \cb{monopolar symmetry} at the bottom, respectively top, wall, we conclude that the symmetry is not spontaneously broken but induced by the gravity and the density mismatch between the water droplet and the surrounding oil phase. 

\section{Conclusion}
In summary, in this work, we presented first-of-a-kind measurements of the 3D flow field produced by a swimming water droplet using PIV method and confocal microscopy in two different configurations, namely that of a droplet swimming parallel and close to a single confining boundary (bottom wall), and that of a doubly-confined droplet swimming between and close to two parallel walls.
In the one wall case, a simplified description of the swimming droplet was proposed as the superposition of (i) the few first axisymmetric viscous and potential singularities (and their image system), accounting for the swimming motion along the wall and (ii) a vertical point force and source dipole (and their image system), accounting for the pumping flow resulting from the top-down asymmetric modification of the chemical environment of the droplet by the confining wall. This model was observed to provide a very good description of the flow field around the droplet, with the swimming-induced singularities (i.e. parallel to the wall) able to reproduce accurately the quadrupolar and dipolar symmetries of the flow field, while the normal singularities associated with the pumping flow was further able to capture the \cb{monopolar symmetry signature of the flow field generated by the droplet}. Adding this second contribution was shown to be critical to effectively capture the dominant flow field far from the droplet, which drives its interactions with its neighbors and environment.

The surprisingly good quality of this description is rooted in the possibility of superimposing the singularities parallel to and orthogonal to the wall, which in turn comes from the linearity of the Stokes equation. This is only valid in the steady state, where the slip velocity at the interface of the swimmer are, by definition, constant. If one were to solve the dynamics of the swimmer, say when approaching the wall with some angle, then the slip velocity at the interface of the swimmer would couple to the hydrodynamics through the concentration field, making the problem truly non linear. Solving for the swimmer trajectory using a quasi-static approximation is a promising route for future work.

Another interesting perspective would be to consider interactions between two confined swimming droplets. One infers from the present results that the long distance interaction between two droplets swimming at the wall is dominated by the \cb{monopolar symmetry}, which is attractive. When the interacting droplets get closer, the quadrupolar symmetry eventually dominates so that, at short enough distance, one recovers the usual pusher/puller qualification and resulting dynamics. 

Finally, in the case of confinement between two walls, we observe an asymmetry with respect to the median plane, and again a monopolar symmetry, which indicates that the gravity can not be neglected, as one could have thought at first sight. The confinement, in this case, imposes that the flows pumped by the swimmer must be evacuated laterally, inducing a recirculation at the scale of the droplet.

\section*{Acknowledgment}
The authors thank Sandra Lerouge for her help in using the PIV method and enlightening discussions.
Charlotte de Blois was sponsored by a doctoral fellowship from the Ecole Doctoral Physique en Ile de France. This work was also supported by the European Research Council (ERC) under the European Union's Horizon 2020 research and innovation program (Grant Agreement No. 714027 to S.M.).

\appendix
\section{\label{App:Meth} Methods : PIV measurement and Data representation}
The PIV analysis is performed using the PIV\_lab~\cite{Thielicke_2014} code on \copyright Matlab. After pre-processing the images with a Wiener filter (window size of $3\times 3$ pixels), we use two successive paths of integration using correlation areas of first $50\times 50$ pixels and then $25\times 25$ pixels, with a sliding step of $50$\% of the correlation area. Doing so we obtain the velocity components $U_x(x,y,z,t)$ and $U_y(x,y,z,t)$ with a spatial resolution of $8\mu$m (around 8\% of the droplet radius) for each pair of successive images. Note that the velocity component $U_z$ is certainly non zero but is not measured here. Note that all distances are made dimensionless using the droplet radius, and all velocities using the droplet velocity. Let us end with stating some limitations of the experimental method. In planes other than the median plane, the quality of the imaging is altered when the incoming light goes through the droplet. As a result the PIV cannot be performed inside the vertical cylinder tangent to the droplet in the median plane. Also, for distances larger than five droplet radius, the signal to noise ratio is too small to extract reliable velocity fields.\\

The first step of the analysis is, for each $z$, to average over time the successive flow fields $U_x$ and $U_y$. These flow fields are dominated by a dipolar symmetry, which is used to extract the position of the droplet center and the direction of its instantaneous displacement and an estimate of it speed $U_0$. We then apply a translation and a rotation to superimpose all successive flow fields at each $z$ and obtain their temporal average $u_x(x,y,z)$ and $u_y(x,y,z)$. \cb{From now on, we choose the droplet radius $\frac{a}{2}$ as the unit of space, and $\frac{a}{2U_0}$ as the unit of time. }
In the presence of a wall, the flow around an axisymmetric swimmer is not axisymmetric and an appropriate base to describe the flow field is the cylindrical base $(\rho,\theta,z)$, defined on Fig.~\ref{fig:1W_3D}.\\  

For a steady and linear motion the flow field conserves a planar symmetry with respect to the plane $\theta=0$ defined by the normal to the wall and the direction of motion.  This parity symmetry allows us to decompose the radial $u_\rho$ and azimuthal $u_\theta$ components of the dimensionless velocity field in each plane, using the basis of Legendre and associated Legendre polynomials of the first kind $L_n(\nu)$ and $L^1_n(\nu)=-\sqrt{1-\nu^2}L_n'(\nu)$:
\begin{align}
u_\rho(\rho,\theta,z)&=\sum_{n=1} \phi^n_\rho(\rho,z) L_n(\nu)  \\
u_\theta(\rho,\theta,z)&=\sum_{n=1} \phi^n_\theta(\rho,z) L_n^1(\nu) 
\label{eq:Legendredecomposition}
\end{align}
where $\nu = cos(\theta)$.
Each Legendre polynomial describes an azimuthal symmetry of the flow in the planes parallel to the wall:  $L_0(\nu)=1$ describes the \cb{monopolar symmetry, $L_1(\nu)=\nu$ the dipolar symmetry, and $L_2=\frac{1}{2}(3\nu^2-1)$ the quadrupolar symmetry}. Since the Legendre polynomials form an orthogonal basis, the amplitudes $\phi^n_{\rho,\theta}(\rho,z)$ are obtained from the the projections:
\begin{align}
\phi^n_\rho(\rho,z)&=\frac{2 n+1}{2}\int_{-1}^1 u_\rho(\rho,\theta,z) L_n(\nu) d\nu,\\
\phi^n_\theta(\rho,z)&=\frac{2 n+1}{2 n (n+1)}\int_{-1}^1 u_\theta(\rho,\theta,z) L_n^1(\nu) d\nu.
\label{eq:projection}
\end{align}
For each $z$, $\phi^n_{\rho,\theta}(\rho,z)$ describes the radial dependency of the $n^{th}$ azimutal symmetry of the velocity $u_{\rho,\theta}$. These amplitudes form the output of our experimental measurements and are represented in the main text on Fig.~\ref{fig:1W}, respectively Fig.~\ref{fig:2W}, in the case of a droplet swimming above one wall, respectively between two walls. 

\section{\label{App:Theo0W} Cylindrical representation of a model swimmer moving in an unbounded fluid}
The flow field around an axisymmetric swimmer can be computed exactly, solving the Stokes equation~\cite{Blake_1971,Lighthill_1952}, for any given slip velocity $\ub_s$ at the interface. However, in agreement with our experimental observations, we limit our description to the terms with azimuthal symmetries up to the quadrupolar order. Recalling that a swimmer is force-free (no external force), and source-free (no net flux production at the interface), the swimmer is modeled by a \cb{stokes dipole}, responsible for the leading order in $1/r$ of the flow field $\ub_{fd}(\rb)\sim 1/r^2$, a source dipole $\ub_{sd}(\rb)\sim 1/r^3$ resulting from the finite size of the swimmer and a source quadrupole, that ensures the absence of normal flux at the interface $\ub_{sq}(\rb)\sim 1/r^4$.  This set of singularities corresponds effectively to the first two modes of the commonly-used squirmer model, which, in an unbounded geometry, generates the flow field:
\begin{equation}
\ub(\rb) = \lambda \ub_{sd}(\rb) + \zeta \ub_{sq}(\rb) + \kappa \ub_{fd}(\rb),
\end{equation}
where the dimensionless coefficients $\lambda=\frac{1}{2}, \kappa=-3\zeta$ are set by the boundary condition $\ub(r=1) = \eb + \ub_s$, with $\eb$ the unit vector pointing in the direction of the swimming motion.
The expressions of the singularities $\ub_{sd}(\rb), \ub_{sq}(\rb)$ and $\ub_{fd}(\rb)$ are obtained from the gradients of the point source $\ub_s(\rb)$ and point force $\ub_f(\rb)$ singularities:
\begin{align}
\ub_s(\rb)&=\frac{\rb}{r^3},\\
\ub_f(\rb)&=\left(\frac{\eb}{r}+\frac{(\eb\cdot\rb)\rb}{r^3}\right),\\
\ub_{sd}(\rb)&=-\left[\nabla\ub_s\right]\cdot\eb=\left(\frac{3(\rb\cdot\eb)\rb}{r^5}-\frac{\eb}{r^3}\right),\label{eq:0W-vect-1}\\
\ub_{sq}(\rb)&=-\left[\nabla\ub_{sd}\right]\cdot\eb=3\left[\frac{5(\rb\cdot\eb)^2\rb}{r^7}-\frac{\rb}{r^5}-\frac{2(\rb\cdot\eb)\eb}{r^5}\right],\\
\ub_{fd}(\rb)&=-\left[\nabla\ub_f\right]\cdot\eb=-\left(\frac{\rb}{r^3}-\frac{3(\rb\cdot\eb)^2\rb}{r^5}\right).
\label{eq:0W-vect-2}
\end{align}

We then write the singularities in the cylindrical system of coordinates $(\rho,\theta,z)$, using $\eb=\nu \eb_\rho -\sqrt{1-\nu^2} \eb_\theta$, $\rb=\rho \eb_\rho + z \eb_z$ and $\rb\cdot\eb=\rho \nu$: 

\begin{align}
    u_{sd,\rho}(\rho,\nu)&=\left(\frac{3\rho^2}{r^5}-\frac{1}{r^3} \right) L_1(\nu) , &
    u_{sd,\theta}(\rho,\nu)&=-\frac{1}{r^3} L_1^1(\nu) ,\\
    u_{sq,\rho}(\rho,\nu)&=\left(\frac{10\rho^3}{r^7}-\frac{4\rho}{r^5}\right) L_2(\nu)  + 5 \left(\frac{\rho^3}{r^7}-\frac{\rho}{r^5}\right) L_0(\nu) , &
    u_{sq,\theta}(\rho,\nu)&= - \frac{2\rho}{r^5} L_2^1(\nu) ,\\
    u_{fd,\rho}(\rho,\nu)&=\frac{2\rho^3}{r^5} L_2(\nu)-  \left(\frac{\rho}{r^3} - \frac{\rho^3}{r^5}\right) L_0(\nu), &
    u_{fd,\theta}(\rho,\nu)&=0.
\end{align}
Note that in the median plane, in which the cylindrical and the more usual spherical coordinate are identical, ($z=0$ and $r=\rho$), \cb{the source quadrupole and the stokes dipole contribute only to the quadrupolar symmetry of the flow}. This is not the case out of the median plane ($z\neq0$), where they also generate a \cb{monopolar symmetry} in the radial velocity component $u_\rho(\rho,\nu)$. This is only reflecting the choice of coordinate system as the three dimensional flow field is inherently the purely dipolar and quadrupolar axi-symmetric squirmer flow field.

Finally, we obtain the amplitude of the monopolar, dipolar and quadrupolar \cb{components}, projecting onto the Legendre polynomials, Eq.~\eqref{eq:projection}: 
\begin{align}
\label{eq:0W-amp-1}
\phi^0_\rho(\rho)&= 5 \zeta \left(\frac{\rho^3}{r^7}-\frac{\rho}{r^5}\right)- \kappa \left(\frac{\rho}{r^3} - \frac{\rho^3}{r^5}\right) ,\\
\phi^1_\rho(\rho)&=\lambda \left(\frac{3\rho^2}{r^5}-\frac{1}{r^3}\right)   ,\\
\phi^2_\rho(\rho)&=\zeta \left(\frac{10\rho^3}{r^7}-\frac{4\rho}{r^5}\right) +  \frac{2\kappa\rho^3}{r^5},\\ 
\phi^1_\theta(\rho)&=-\frac{\lambda}{r^3}   ,\\
\phi^2_\theta(\rho)&=-\frac{2 \zeta \rho}{r^5}  .
\label{eq:0W-amp-2}
\end{align}

\begin{figure}[h!]
\includegraphics[width=0.7\columnwidth ]{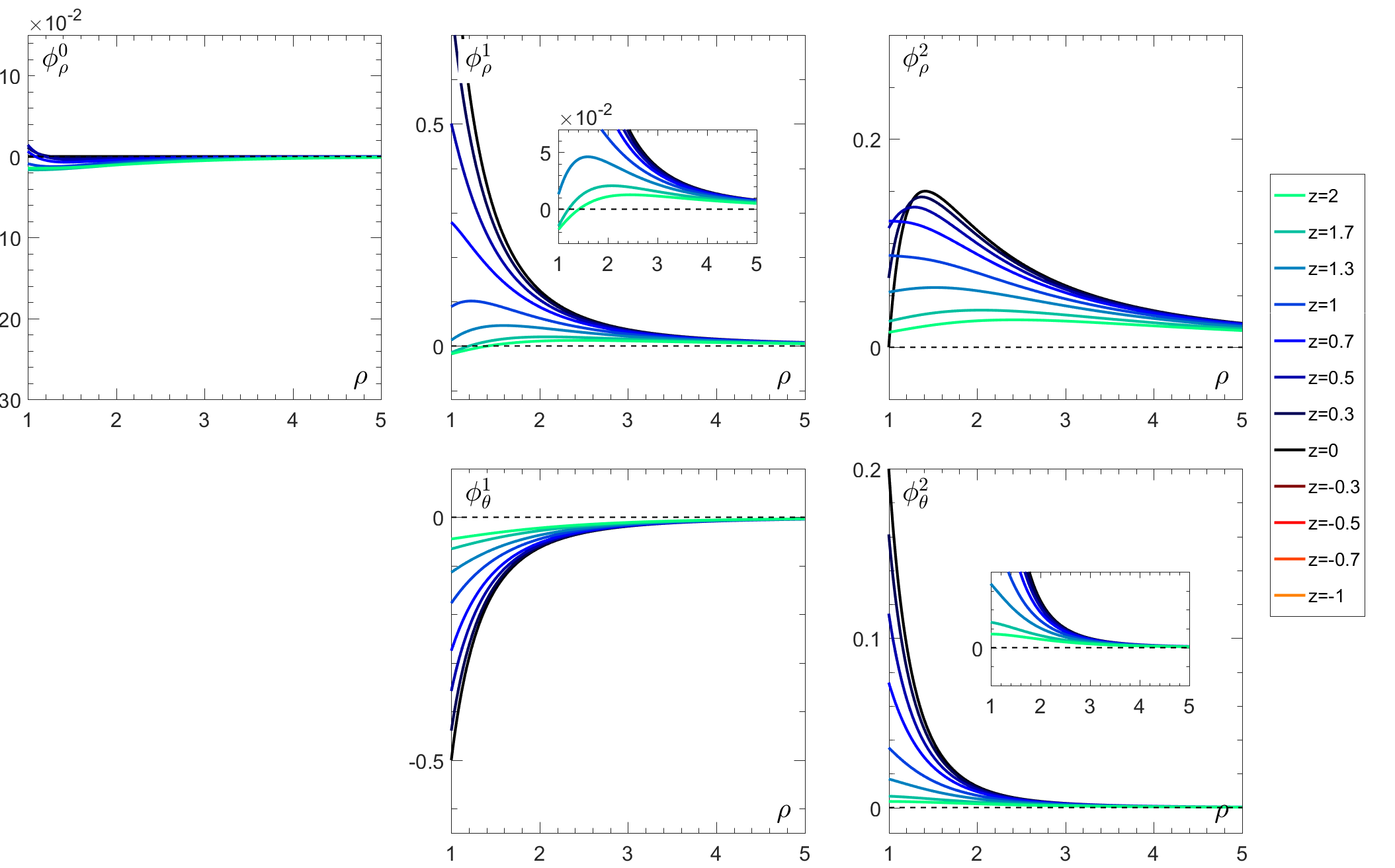}
\caption{{\bf Flow field created by a squirmer in the unbounded case}: amplitude of the monopolar (left), dipolar (middle) and  quadrupolar (right) \cb{components of the velocities} $u_\rho$ (top) and $u_\theta$ (bottom), as given by Eqs.~\eqref{eq:0W-amp-1}--\eqref{eq:0W-amp-2}.  The z-planes are color coded according to the legend on the right; same color code as in the main text. The $z>0$ and $z<0$ profiles superimpose. The coefficients $\lambda=\frac{1}{2}$ and $\zeta=-\frac{1}{3}\kappa$ are given by the boundary condition at the interface of the droplet. $\kappa$ is free and arbitrarily set to $\kappa=0.3$.}
\label{fig:0W_theo}
\end{figure}
Figure~\ref{fig:0W_theo} displays the radial dependence of these amplitude in different z-plane (color code). The black curves correspond to the median plane ($z=0$). 

\section{\label{App:Theo1W} Cylindrical representation of a model swimmer moving parallel to a wall}

As described in the main text, we start again with the same set of singularities as for the standard squirmer model:
\begin{equation}
\ub^*(\rb) =  \lambda \ub^*_{sd}(\rb) + \zeta \ub^*_{sq}(\rb) + \kappa \ub^*_{fd}(\rb),\label{eq:sing_wall}
\end{equation}
The effects of a nearby no-slip wall is taken into account by introducing the image system of each singularity used to describe the swimmer~\cite{Lee_1979}, so that the total flow field satisfies the no slip condition at the wall $\ub^*(z=0)=0$. Such image systems can be computed for each singularity, at any distance and a priori for any angle with the wall, although a particular care should be taken to compute the image of singularities such as the stokes dipole, etc...~\cite{Blake_1974,Spagnolie_2012}. 
We still use a set of axes centered on the sphere's center. $\eb_z$ is a unit vector along the vertical axis and $h$ is the distance between the center of the swimmer and the wall. Image singularities are thus positioned at a point $\Xb=-h\eb_z$ below the wall, and we note $\rb$ the position of the observation point (where the flow is evaluated) with respect to the sphere's center and $\Rb=\rb+2h\eb_z$ the position of the same point with respect to the position of the image system. For each singularity, which leads to a velocity field $\ub_i$ in unbounded flow, we denote by $\ub^*_i$ the corresponding flow field near a wall (i.e. including both the original singularity and the effect of its image system). This flow field is obtained for a point source or point force as~\cite{Blake_1974}:
\begin{align}
\label{eq:1W-us}
\ub_s^*(\rb)=&\overbrace{\frac{\rb}{r^3}}^\textrm{Original source}+\overbrace{\frac{\Rb}{R^3}}^\textrm{Image source}-\overbrace{2\left(\frac{\Rb}{R^3}-\frac{3\Rb(\Rb\cdot\eb_z)^2}{R^5}\right)}^\textrm{stresslet}+\overbrace{2h\left(\frac{\eb_z}{R^3}-\frac{3(\Rb\cdot\eb_z)\Rb}{R^5}\right)}^\textrm{source dipole},\\
\ub_f^*(\rb)=&\overbrace{\frac{\eb}{r}+\frac{(\eb\cdot\rb)\rb}{r^3}}^\textrm{Original stokeslet}-\overbrace{\frac{\eb}{R}-\frac{(\eb\cdot\Rb)\Rb}{R^3}}^\textrm{Image Stokeslet}+\overbrace{\frac{2h^2\eb}{R^3}-\frac{6h^2(\eb\cdot\Rb)\Rb}{R^5}}^\textrm{source dipole}+\overbrace{\frac{2h(\eb\cdot\Rb)\eb_z}{R^3}+\frac{6h(\Rb\cdot\eb_z)(\eb\cdot\Rb)\Rb}{R^5}-\frac{2h(\eb_z\cdot\Rb)\eb}{R^3}}^\textrm{stresslet}.
\label{eq:1W-uf}
\end{align}
In our model, the singularities used in Eq.~\eqref{eq:sing_wall} are all parallel to the wall. The flow field generated by higher order singularities is thus obtained by taking successive gradients of the flow field generated by a point force or point source singularity. As an example, the flow field $\ub^*_{sd}$, generated by the source dipole singularity $\ub_{sd}$ in the presence of a no-slip wall, is obtained by taking the gradient of the flow field $\ub^*_s$, generated by a source monopole singularity and project it on $\eb$. The methods applies iteratively to obtain the flow fields generated by higher-order singularities:
\begin{align}
\label{eq:1W-usd}
\ub_{sd}^*(\rb)=&-\left[\nabla\ub_s^*\right]\cdot\eb\nonumber\\
=&-\frac{\eb}{r^3}+\frac{3(\eb\cdot\rb)\rb}{r^5}+\frac{\eb}{R^3}-\frac{3(\eb\cdot\Rb)\Rb}{R^5}-6(\Rb\cdot\eb_z)(\Rb\cdot\eb_z-h)\left(\frac{\eb}{R^5}-\frac{5(\Rb\cdot\eb)\Rb}{R^7}\right)+\frac{6h(\Rb\cdot\eb)\eb_z}{R^5},\\
\label{eq:1W-usq}
\ub_{sq}^*(\rb)=&-\left[\nabla\ub_{sd}^*\right]\cdot\eb\nonumber\\
=&\frac{15(\rb\cdot\eb)^2\rb}{r^7}-\frac{3\rb}{r^5}-\frac{6(\rb\cdot\eb)\eb}{r^5}-\frac{15(\Rb\cdot\eb)^2\Rb}{R^7}+\frac{3\Rb}{R^5}+\frac{6(\Rb\cdot\eb)\eb}{R^5}\nonumber\\
&+30h(\Rb\cdot\eb_z-h)(\Rb\cdot\eb_z)\left(\frac{7(\Rb\cdot\eb)^2\Rb}{R^9}-\frac{2(\Rb\cdot\eb)\eb}{R^7}-\frac{\Rb}{R^7}\right) -\frac{6h\eb_z}{R^5}+\frac{30h(\Rb\cdot\eb)^2\eb_z}{R^7},\\
\label{eq:1W-ufd}
\ub_{fd}^*(\rb)=&-\left[\nabla\ub_f^*\right]\cdot\eb\nonumber\\
=&-\frac{\rb}{r^3}+\frac{3(\eb\cdot\rb)^2\rb}{r^5}+\frac{\Rb}{R^3}-\frac{3(\eb\cdot\Rb)^2\Rb}{R^5}-\frac{6h(\Rb\cdot\eb_z-h)\Rb}{R^5}-\frac{12h(\Rb\cdot\eb_z-h)(\eb\cdot\Rb)\eb}{R^5}\nonumber\\
&+\frac{30h(\eb\cdot\Rb)^2(\Rb\cdot\eb_z-h)\Rb}{R^7}-\frac{2h\eb_z}{R^3}+\frac{6h(\Rb\cdot\eb)^2\eb_z}{R^5}.
\end{align}

These velocity fields are then projected along the axes of the cylindrical coordinate systems, using that $\Rb = (2h+z) \eb_z + \rho\,\eb_{\rho}$, $\Rb\cdot \eb = \rb \cdot \eb = \rho \nu$ and $\Rb \cdot \eb_z = 2h+z$ :

\begin{align}
u_{sd,\rho}^*(\rho,\nu)=& \left[-\frac{1}{r^3}+\frac{1}{R^3}-3\rho^2\left(\frac{1}{R^5}-\frac{1}{r^5}\right)-6(h+z)(2h+z)\left(\frac{1}{R^5}-\frac{5\rho^2}{R^7}\right)\right]L_1(\nu),\\
u_{sd,\theta}^*(\rho,\nu)=&\left(-\frac{1}{r^3}+\frac{1}{R^3}-\frac{6(2h+z)(h+z)}{R^5}\right)L_1^1(\nu) ,
\end{align}
and
\begin{align}
u_{sq,\rho}^*(\rho,\nu)=&\left[\frac{10\rho^3}{r^7}-\frac{4\rho}{r^5}-\frac{10\rho^3}{R^7}+\frac{4\rho}{R^5}+20h(z+h)(z+2h)\left(\frac{7\rho^3}{R^9}-\frac{2\rho}{R^7}\right)\right] L_2(\nu)\nonumber\\
&+\left[\frac{5\rho^3}{r^7}-\frac{5\rho}{r^5}-\frac{5\rho^3}{R^7}+\frac{5\rho}{R^5}+10h(z+h)(z+2h)\left(\frac{7\rho^3}{R^9}-\frac{5\rho}{R^7}\right)\right]L_0(\nu),\\
u_{sq,\theta}^*(\rho,\nu)=&\left(-\frac{2\rho}{r^5}+\frac{2\rho}{R^5}-\frac{20h(z+h)(z+2h)\rho}{R^7}\right) L_2^1(\nu),\\
u_{fd,\rho}^*(\rho,\nu)=&\left(\frac{2\rho^3}{r^5}-\frac{2\rho^3}{R^5}-\frac{8h(z+h)\rho}{R^5}+\frac{20h\rho^3(z+h)}{R^7}\right) L_2(\nu)\nonumber\\
&+\left[-\frac{\rho}{r^3}+\frac{\rho^3}{r^5}+\frac{\rho}{R^3}-\frac{\rho^3}{R^5}-10h(z+h) \left(\frac{\rho}{R^5}-\frac{\rho^3}{R^7}\right)\right]L_0(\nu),\\
u_{fd,\theta}^*(\rho,\nu)=&\,\,-4h(z+h)\frac{\rho}{R^5}L_2^1(\nu).
\end{align}

Finally, we obtain the amplitudes by projection onto the Legendre polynomials: 
\begin{align}
\label{eq:1W-amp-1}
\phi^{0*}_\rho(\rho,z)=&\zeta \left[\frac{5\rho^3}{r^7}-\frac{5\rho}{r^5}-\frac{5\rho^3}{R^7}+\frac{5\rho}{R^5}+10h(z+h)(z+2h)\left(\frac{7\rho^3}{R^9}-\frac{5\rho}{R^7}\right)\right]\nonumber\\
& + \kappa \left[-\frac{\rho}{r^3}+\frac{\rho^3}{r^5}+\frac{\rho}{R^3}-\frac{\rho^3}{R^5}-10h(z+h) \left(\frac{\rho}{R^5}-\frac{\rho^3}{R^7}\right)\right] ,\\
\phi^{1*}_\rho(\rho,z)=& \lambda \left[-\frac{1}{r^3}+\frac{1}{R^3}-3\rho^2\left(\frac{1}{R^5}+\frac{1}{r^5}\right)-6(h+z)(2h+z)\left(\frac{1}{R^5}-\frac{5\rho^2}{R^7}\right)\right],\\
\phi^{2*}_\rho(\rho,z)=&\zeta \left[\frac{10\rho^3}{r^7}-\frac{4\rho}{r^5}-\frac{10\rho^3}{R^7}+\frac{4\rho}{R^5}+20h(z+h)(z+2h)\left(\frac{7\rho^3}{R^9}-\frac{2\rho}{R^7}\right)\right]\nonumber\\
& + \kappa \left(\frac{2\rho^3}{r^5}-\frac{2\rho^3}{R^5}-\frac{8h(z+h)\rho}{R^5}+\frac{20h\rho^3(z+h)}{R^7}\right)  ,\\ 
\phi^{1*}_\theta(\rho,z)=&\lambda\left(-\frac{1}{r^3}+\frac{1}{R^3}-\frac{6(2h+z)(h+z)}{R^5}\right) ,\\
\phi^{2*}_\theta(\rho,z)=&\zeta \left(-\frac{2\rho}{r^5}+\frac{2\rho}{R^5}-\frac{20h(z+h)(z+2h)\rho}{R^7}\right)  \nonumber\\
&-   \frac{4\kappa\rho h(z+h)}{R^5} .
\label{eq:1W-amp-2}
\end{align}

\section{\label{App:TheoPerp} Cylindrical representation of a perpendicular point force and source dipole near a wall}
No singularity parallel to the swimming direction can give rise to a \cb{monopolar symmetry} of the flow in an unbounded fluid when the swimmer is source-free, as is the case here for the water droplet (the source flow associated with the water leaving the droplet in swollen micelles is indeed negligible on the time scale of the measurements). However, one may notice that a \cb{dipolar flow observed in a plane perpendicular to its axis of symmetry includes a monopolar component. Thus a source dipole or a point force singularities - that give rise to a dipolar symmetry in planes parallel to its axis of symmetry - would give rise to a monopolar symmetry in planes parallel to the swimming direction if perpendicular to this swimming direction.}

Motivated by this observation, we therefore consider the flow field generated by a point force and source dipole normal to the wall and their associated images, which write in vector form :

\begin{align}
\ub_{pf,\perp}^*(\rb)=&\frac{\eb_z}{r}+\frac{(\eb_z\cdot\rb)\rb}{r^3}-\frac{\eb_z}{R}-\frac{(\eb_z\cdot\Rb)\Rb}{R^3}-\frac{2h^2\eb_z}{R^3}+\frac{6h^2(\Rb\cdot\eb_z)\Rb}{R^5} +\frac{2h\Rb}{R^3}-\frac{6h(\Rb\cdot\eb_z)^2\Rb}{R^5},\\
\ub_{sd,\perp}^*(\rb)=&
-\frac{\eb_z}{r^3}+\frac{3(\rb\cdot\eb_z)\rb}{r^5}+\frac{3\eb_z}{r^3}-\frac{9(\rb\cdot\eb_z)\rb}{r^5}+\frac{-2\eb_z}{R^3}+\frac{18(\eb_z\cdot\Rb)\Rb}{R^5}+\frac{6(\Rb\cdot\eb_z)^2\eb_z}{R^5}-\frac{30(\eb_z\cdot\Rb)^3\Rb}{R^7}\nonumber\\
&-\frac{12h(\eb_z\cdot\Rb)\eb_z}{R^5}-6\frac{\Rb}{R^5}-\frac{10h(\eb_z\cdot\Rb)^2\Rb}{R^7},
\end{align}
and in cylindrical coordinates: 
\begin{align}
u^*_{pf,\perp,\rho}(\rho,\nu)=&\left(\frac{\rho z}{r^3}-\frac{\rho z}{R^3}-\frac{6h(z+h)(z+2h)\rho}{R^5}\right)L_0(\nu),\\
u^*_{pf,\perp,\theta}(\rho,\nu)=&\,\,0,\\
u^*_{sd,\perp,\rho}(\rho,\nu)=&-\left[\nabla\ub_s^*\right]\cdot\eb\nonumber\left[\frac{3z\rho}{r^5}-\frac{3z\rho}{R^5}+3(h+z)\left(\frac{4\rho}{R^5}-\frac{10(2h+z)^2}{R^7}\right)\right]L_0(\nu),\\
u^*_{sd,\perp,\theta}(\rho,\nu)=&0.
\end{align}

\cb{The angular dependence} of these two singularities is purely monopolar in this cylindrical base. If we note $\epsilon$ the coefficients of the perpendicular point force, and $\sigma$ the amplitude of the perpendicular source dipole, then the amplitude of the monopolar symmetry to the flow field, taking into account all previous singularities close to a wall is : 

\begin{align}
\phi_0^*(\rho,z)=&\epsilon\left(\frac{\rho z}{r^3}-\frac{\rho z}{R^3}-\frac{6h(z+h)(z+2h)\rho}{R^5}\right)\nonumber\\
 &+\sigma \left[\frac{3z\rho}{r^5}-\frac{3z\rho}{R^5}+3(h+z)\left(\frac{4\rho}{R^5}-\frac{10(2h+z)^2}{R^7}\right)\right]\nonumber\\
&+\zeta \left[\frac{5\rho^3}{r^7}-\frac{5\rho}{r^5}-\frac{5\rho^3}{R^7}+\frac{5\rho}{R^5}+10h(z+h)(z+2h)\left(\frac{7\rho^3}{R^9}-\frac{5\rho}{R^7}\right)\right]\nonumber\\
& + \kappa \left[\frac{\rho}{r^3}-\frac{\rho^3}{r^5}-\frac{\rho}{R^3}+\frac{\rho^3}{R^5}+10h(z+h) \left(\frac{\rho}{R^5}-\frac{\rho^3}{R^7}\right)\right].
\end{align}

\bibliography{Swimmer}

\end{document}